\documentclass[paper]{JHEP3}

\title{Wilson Loops for a quark anti-quark pair in D3-brane space}

\author{Henrique Boschi-Filho\\
Instituto de F\'{\i}sica, Universidade Federal do Rio de Janeiro, 
Caixa Postal 68528, RJ 21945-970 -- Brazil\\ 
E-mail: \email{boschi@if.ufrj.br}}

\author{Nelson R. F. Braga\\
Instituto de F\'{\i}sica, Universidade Federal do Rio de Janeiro, 
Caixa Postal 68528, RJ 21945-970 -- Brazil\\
E-mail: \email{braga@if.ufrj.br}}

\abstract  
{We calculate static Wilson loops for a heavy quark anti-quark pair in 
different positions in the space generated by a large number of coincident D3-branes. 
Simple results are obtained from limiting cases of the geodesic shape.
In particular, quark anti-quark static potentials for flat and AdS spaces 
are reproduced.}

\keywords{ads,dbr}

\preprint{hep-th/0411135}

\begin{document}

\section{Introduction}
It was shown by Maldacena that large N superconformal gauge theories have a dual
description in terms of string theory in AdS space\cite{Maldacena:1997re}. 
Soon after, this proposal was used to calculate Wilson loops for gauge theories 
from the corresponding dual geometry\cite{RY,MaldaPRL}. 
This allows a computation of the interaction energy 
associated with gauge fields leading to a geometric criterion for 
confinement\footnote{The boundary conditions 
of this problem have been discussed in \cite{RY}.}. 
For instance, the energy of a quark anti-quark pair in large N 
superconformal ${\cal N} = 4$ Yang-Mills theory is obtained from the Wilson 
loop of the corresponding string in AdS space. 
For this space the energy has a non-confining Coulomb like behaviour,
as expected for a conformal field theory.
This approach was applied to  multicentre and rotating 
branes\cite{Minahan:1998xq,Brandhuber:1999jr} and 
many other spaces and models\cite{Greensite:1998bp,Brandhuber:1998er,Kinar:1998vq,
Sonnenschein:2000qm}. 
More recently this approach to Wilson loops has been used in the search for
confining geometries\cite{Maldacena:2000yy,Klebanov:2000nc,Klebanov:2000hb,
Polchinski:2000uf}
and their related properties\cite{Janik:2000aj,Janik:2001sc,Takahashi:2002cc,
Karch:2002xe,PandoZayas:2003yb,Bigazzi:2004ze}.
Wilson loops have also been applied to discuss scattering 
amplitudes\cite{Rho:1999jm,Janik:1999zk} and radiation\cite{Sin:2004yx}
using AdS/CFT correspondence.

Here we will study Wilson lines in the space generated by a large number of
coincident D3-branes. This space is a particular case of the ones considered 
in\cite{Minahan:1998xq,Brandhuber:1999jr} but here we will consider the quarks 
located at different positions rather than considering them always at infinity.
This is interesting because the curvature of the D3-brane space is not constant.
Near the branes this space is asymptotically AdS while far from the branes
the metric is asymptotically flat. So the behaviour of the geodesics depends 
on the position of the quarks.  We obtain general relations for the energy of
the quark pair in the D3-brane space. Then considering limiting 
cases we obtain interesting simple results for this energy, 
including those corresponding to flat and AdS spaces.

The area of the Wilson loop for a stationary configuration is the product of the 
time interval times the string length. Then the energy of a quark anti-quark pair  
is proportional to this area. In order to obtain this energy we
start describing the string dynamics by the Nambu-Goto action 

\begin{equation}
S \,=\,{1\over 2\pi}\, \int d\sigma d\tau \sqrt{ det \big( g_{MN} \partial_\alpha X^M 
\partial_\beta X^N \Big)\,\,}\,
\end{equation}

\noindent where we are setting the string scale $\alpha^\prime\,=\,1$. 
 The geometry of interest corresponds to a ten-dimensional metric of the form
\begin{equation}
ds^2 \,=\, -g_{00}(r) dt^2 + g_{ii} (r) dx^i dx^i + g_{rr} (r) dr^2 +
d {\tilde s}^2 
\end{equation}
\noindent where $i= 1,2,3 $ and $d {\tilde s}^2 $ represents five extra transverse
directions  that will not be relevant in our discussion. 

 We consider an infinitely heavy quark anti-quark pair located at  $r = r_1$
so that their configuration is always stationary.
For simplicity we can take them as sitting on one of the  $x^i$ axis separated 
by a coordinate distance $L$. The string connecting these quarks represents the geodesic
in this space and reaches a minimum value $r = r_0$. This minimum value
is determined from the equations of motion in terms of $r_1$ and $L$.
Following \cite{Kinar:1998vq} we write a relation between these parameters
as 

\begin{equation}
\label{LGgeral}
L \,=\, \int^{L/2}_{-L/2} dx \,=\, 2 \int_{r_0}^{r_1} 
\Big( {dr\over dx}\Big)^{-1} dr \,=\, 
2 \int^{r_1}_{r_0} { g(r) \over f (r) } { f (r_0)\over \sqrt{ f^2 (r) - f^2 (r_0 )}}
\,dr 
\end{equation} 

\noindent where 
\begin{eqnarray}
f^2 (r) &=& (2\pi )^{-2} g_{00} (r)\,\, g_{ii} (r)\nonumber\\
g^2 (r) &=& (2\pi )^{-2} g_{00} (r) \,\, g_{rr} (r)\,.\label{fg}
\end{eqnarray}

Then the corresponding interaction energy of the quark anti-quark pair is given by

\begin{equation}
\label{EGgeral}
E \,=\, 2 \int^{r_1}_{r_0} { g(r) f (r )\over \sqrt{ f^2 (r) - f^2 (r_0 )}}
\,dr \,-\,2 m_q
\end{equation}

\noindent where $m_q = m_q (r_1) $ is the energy of each non interacting quark.

\section{D3-brane space} 

The invariant measure for the ten 
dimensional geometry generated by a large number $N$ of coincident D3-branes 
can be written as\cite{Horowitz:1991cd,Gubser:1998bc}

\begin{equation}
\label{branemetric}
ds^2 \,=\, \Big( 1 + {R^4\over r^4} \Big)^{-1/2} ( -dt^2 + d{\vec x}^2 ) +  
\Big( 1 + {R^4\over r^4} \Big)^{1/2} (dr^2 + r^2 d\Omega^2_5 )
\end{equation}

\noindent where $R^4 \,=\, 4 \pi g N$.

 
Using this metric in eq. (\ref{LGgeral}) and (\ref{EGgeral}) we obtain 

\begin{equation}
\label{Lgeral}
L \,=\, {2 r_0^3 \over R^2} \, I_1 ( y_1 ) \,+\,
{2 R^2 \over r_0}\,  I_2 ( y_1 ) 
\end{equation}

\begin{equation}
\label{Egeral}
E \,=\,{2 r_0   \sqrt{ r_0^4 + R^4 }\over 2\pi\,R^2} \,I_1 ( y_1 )
\,-\, 2 m_q
\end{equation} 

\noindent where we defined $ y_1 \equiv r_1/r_0\,$ and

\begin{equation}
\label{I1geral}
I_1 ( y_1 ) \,=\, \int_1^{y_1} {y^2 dy \over \sqrt{ y^4 -1}}
\end{equation}

\begin{equation}
\label{I2geral}
I_2 ( y_1 ) \,=\, \int_1^{y_1} {dy \over y^2 \sqrt{ y^4 -1}}
\end{equation}

\noindent This implies that the energy and the quark coordinate separation 
are related by
\begin{equation}
\label{Egeral2}
E \,=\,{ \sqrt{ r_0^4 + R^4 }\over 2\pi \,r_0^2} \left( L \,-\,
2 {R ^2 \over r_0} \, I_2 \right)
\,-\, 2 m_q
\end{equation}

The dependence of this energy on $L$ indicates if the theory is confining or not. 
Note that $r_0$ depends on $L$ because of equation (\ref{Lgeral})
so that this relation may lead to different confining behaviours.
To determine these behaviour one must evaluate these elliptic integrals that 
depend on the quark position.
We are going to consider some limiting cases of the parameter
$y_1 =  r_1 / r_0\,$ that represents the shape of the geodesics.
These simple particular cases correspond to interesting physical situations.

\section{ First case:  highly curved geodesics }

This situation is defined by the condition $y_1 \equiv r_1 / r_0\, >> 1$. 
We will find different subcases corresponding to the minimum of the geodesic 
$r_0$ close to or far from the branes.

In this case it is convenient to consider the series expansion of
the integrals (\ref{I1geral},\ref{I2geral}):

\begin{equation}
\label{I11}
I_1 ( y_1 ) \,=\, - C_1 + y_1 - {1\over 6 y_1^3} + O(y_1^{-7})
\end{equation}

\begin{equation}
\label{I22}
I_2 ( y_1 ) \,=\, C_2 - {1\over 3 y_1^3} + O(y_1^{-7}) \,,
\end{equation}

\noindent where $C_1$ and $C_2$ are constants. 
In order to calculate these constants we evaluate the integrals 
in the limit $y_1 \rightarrow \infty$ using beta functions. The first 
integral is divergent in this limit but the difference $I_1 - y_1  $  is finite.
So we get 
\begin{equation}
C \equiv C_1 = C_2  = {\pi\sqrt{2\pi}\over \Gamma^2 (1/4)}.
\end{equation}

Inserting in eqs. (\ref{Lgeral}) and (\ref{Egeral}) these
polynomial approximations  we find

\begin{equation}
\label{L1}
L \,=\, {2 r_0^2\over R^2} r_1 \,-\, {2 r_0^3\over R^2} C \,+\, 
{2 R^2\over r_0} C
\end{equation}

\begin{equation}
\label{E1}
E \,=\,{2 \sqrt{ r_0^4 + R^4 }\over 2\pi\, R^2} ( r_1 - C r_0 )
\,-\, 2 m_q
\end{equation} 
\noindent where we disregarded terms of order $O(y_1^{-3})\,$.

Note that eq. (\ref{L1}) can be solved for $r_0$ in terms of the 
position $r_1$ of the quarks, their separation $L$ and the constant $R$.
Using again the fact that $r_1 >> r_0$ this equation reduces to a cubic one

\begin{equation}
r_0^3 \,-\,{L R^2\over 2 r_1} r_0 \,+\, {R^4 C \over r_1}\,=\,0\,\,.
\end{equation} 

\noindent If $ \Delta \equiv {L^3 \over 6^3 r_1} - ({C R\over 2})^2 \,<\, 0$ this 
equation has one negative real and two complex solutions. Then the  physically acceptable
solutions come from $ \Delta \ge 0\,$. The interesting solutions correspond to
$\Delta > 0$ since the case $\Delta = 0$ fixes $L$ in terms of $r_1$.
For $\Delta > 0$ there are two real 
positive solutions  

\begin{equation}
\label{r0+}
r_0^+ \,=\,2 \,\sqrt{{R^2 L\over 6 r_1}}\, \cos \left[ \theta_1 \right]
\end{equation}

\begin{equation}
\label{r0-}
r_0^- \,=\,2\,\sqrt{{R^2 L\over 6 r_1}}\,\sin \left[ \theta_1 \,- { \pi \over 6} \right]
\end{equation}

\noindent where 
$$ \theta_1 \,=\, {1\over 3}\, \cos^{-1} \left( - 3 {R C \over L }
\sqrt{{ 6 r_1 \over L}} \right).$$

\noindent Note that $r_0^+ \,\ge\,r_0^- \,$ and there is also one non physical 
negative solution.   

Note that only one of these two roots minimizes the energy. 
One can see from eq. (\ref{E1}) which of the solutions will be appropriate,
depending on the ratio $r_0/R$.
For the interesting limiting cases $r_0 << R$ and $r_0 >> R $
the solutions that minimize the energy are $r_0^+$ and  
$r_0^-$ respectively. 
Let us now consider these particular subcases. 

\subsection{ Geodesic minimum close to the branes}

In this case, corresponding to $\,r_0 << R$, 
the minimum of the energy is given by $r_0 \,=\,r_0^+$.
Then one could substitute expression (\ref{r0+}) in the energy (\ref{E1}).
However we can get a nicer picture if we use the limit  $\,r_0 << R$   
directly in eqs. (\ref{L1}) and (\ref{E1}) finding

\begin{equation}
L \,=\, {2 R^2\over r_0} C
\end{equation}

\begin{equation}
E \,=\,{ r_1 \over \pi} -  {C r_0\over \pi}  -\, 2 m_q\,.
\end{equation} 

\noindent Identifying $r_1/2\pi\,$ as the energy of each quark and 
using the relation between $L$ and $r_0$ we recover the result of 
\cite{RY,MaldaPRL} for the $AdS$  space

\begin{equation}
\label{EAdS}
E \,= - \, {4 \, C^2 R^2 \over 2\pi\,L}\,.
\end{equation} 

\noindent This result is consistent with the fact that close to the branes 
the D3-brane space is asymptotically AdS. Since this energy 
decreases with $L$ it shows a non confining behaviour. 
Note that the above result is an approximation valid only for $r_1 >> r_0\,$.
So this expression becomes exact if $r_1 \rightarrow \infty\,$.
If one wants to obtain the corrections to this energy in terms of powers of
$r_1$ one should include the negative powers appearing in expansion (\ref{I11})
and also expand the cosine that shows up in (\ref{r0+}).
Although this approach would be valid for any $r_1 > r_0 $, 
it is not convenient in the case $r_1 \sim r_0$. 
We will consider an alternative approach for $r_1 \sim r_0$ in section {\bf 4}.

\subsection{ Geodesic minimum far from the branes}

In this case, corresponding to $ r_0 >> R$,
the minimum of the energy is given by $r_0 \,=\,r_0^-$.
As in the previous case, it is more convenient to consider the approximation
$ r_0 >> R$ directly in eqs. (\ref{L1}) and (\ref{E1}) obtaining 

\begin{equation}
L \,=\, {2 r_0^3\over R^2}\left( {r_1\over r_0}  \,-\,  C  \right)
\end{equation}

\begin{equation}
E \,=\,{L \over 2\pi}\,-\, 2 m_q
\end{equation} 

\noindent This leading order approximation for the energy coincides with 
the flat space case.  This is consistent with the fact that far from the branes
the D3-brane space is asymptotically flat. This energy increases with $L$ so that
it exhibit a confining behaviour.

\section{Second case: almost straight geodesics }

This situation is characterized by the condition $r_1 \sim r_0\,,$ which means 
$ \, y_1 \equiv r_1 / r_0 \,=\, 1 + \epsilon $ with $\epsilon << 1$.
As in the previous case we will also find subcases depending on the position of 
the geodesic minimum $r_0$ far from or close to the branes. 

In this case equations (\ref{I1geral}) and (\ref{I2geral})
reduce to  

\begin{eqnarray} 
I_1 ( 1 + \epsilon ) &=& \sqrt{\epsilon}\Big( 1 + O(\epsilon) \Big)\\
I_2 ( 1 + \epsilon ) &=& \sqrt{\epsilon}\Big( 1 + O(\epsilon) \Big)
\end{eqnarray}

\noindent so that quark coordinate separation (\ref{Lgeral}) and interaction 
energy (\ref{Egeral}) become

\begin{equation}
\label{LA}
L \,=\, 2 \left( { r_0^3 \over R^2}  \,+\,
{ R^2 \over r_0}\,\right)  \sqrt{ \,{r_1\over r_0} - 1\,\,} 
\end{equation}

\begin{equation}
\label{EA}
E \,=\,{ \sqrt{ r_0^4 + R^4 }\over 2\pi\, r_0^2} \left( 
 {2 r_0^3 \over R^2 }\, \sqrt{\, {r_1\over r_0} - 1\,\,} \right)
\,-\, 2 m_q
\end{equation}

In order to obtain simple solutions for this system with interesting physical 
interpretation we consider next the limiting cases of the ratio $r_0/R$ as in 
the previous section.

\subsection{ Quarks far from the branes}
If we put the quarks far from the branes, that means $r_1 >> R$ the condition
$r_1 \sim r_0\,$ implies $r_0 >> R\,,\,$ so that eqs. (\ref{LA}) and (\ref{EA}) 
take the approximate form 

\begin{eqnarray}
L &=& { 2 r_0^3 \over R^2} \sqrt{ {r_1\over r_0} - 1}\\
E &=& {L\over 2\pi}  \,-\, 2 m_q
\end{eqnarray}

\noindent This interaction energy between quarks corresponds to the flat space
case. This is expected since the space felt by the quarks far from the branes  
is approximately flat, as can be seen from the metric (\ref{branemetric}) in the limit
$r >> R$.

\subsection{ Quarks near the branes}
Putting the quarks close to the branes, that means $r_1 << R$ the condition
$r_1 \sim r_0\,$ implies $r_0 << R$ 
and the expressions (\ref{LA}) and (\ref{EA}) are approximated by
 
\begin{equation}
\label{LB}
L \,=\, 2 { R^2 \over r_0}\,  \sqrt{ \,{r_1\over r_0} - 1\,\,} 
\end{equation}

\begin{equation}
\label{EB}
E \,=\, {L \,r_0^2\over 2\pi\,R^2}\,- 2 m_q \,.
\end{equation}

In order to obtain a relation between the energy and the quark separation $L$
one needs to solve equation (\ref{LB}) for $r_0$. 
Using the approximations considered in this case we find

\begin{equation}
\label{sol}
r_0 \,=\, \sqrt[{\displaystyle 3}]{\,{2\over 3}\,}\, \,
r_1 \,\left( 1 - {3 r_1^2 L^2 \over 8 R^4} \right) \,\,.
\end{equation}

\noindent This implies that the energy (\ref{EB}) is approximated by

\begin{equation}
\label{Enear}
E \,=\, \left({2\over 3}\right)^{2/3}\,{L \,r_1^2 \over 2\pi\,R^2}\,\, 
\left( 1 - {3 r_1^2 L^2 \over 8 R^4} \right)^2
  - 2 m_q \,\,.
\end{equation}

\noindent Then the interaction energy between quarks 
is proportional, to leading order, to the coordinate separation 
$L$ times the AdS metric factor $ (r/R)^2 $. 
This can be understood noting that in this case
the quark anti-quark pair is very close to the branes.
So the geodesic is almost a straight line because the transverse
part of the metric in this region as well as its variations against radial
direction are very small.

This result may be surprising if compared with the result 
of references  \cite{RY,MaldaPRL} reproduced in eq. (\ref{EAdS}).
Both cases correspond to quark anti-quark energy configurations in 
AdS space but in section {\bf 3.1} the quarks are 
far from the geodesic minimum while in this section we consider the 
quarks near the geodesic minimum $r_1 \sim r_0 \,$.
One may wonder if the result (\ref{Enear}) above 
could be obtained from (\ref{EAdS}) but as we discussed in the end of 
section {\bf 3.1} the approximations considered there are only valid 
for $r_1 >> r_0$. Both cases correspond to particular approximations  
of equation (\ref{Egeral}) or equivalently (\ref{Egeral2}).

\section{Conclusions}
We calculated the Wilson lines for a quark anti-quark pair 
in the space generated by a large number of D3-branes
with different quark positions.
We have seen that it is possible to recover from the D3-brane space
the Wilson line behaviour corresponding to both AdS and flat spaces
by choosing conveniently the quark position and the geodesic curvature.
It is interesting to note that the D3-brane space exhibits different 
confining behaviours depending on the quark position and geodesic shape.

\section*{Acknowledgments}  
We would like to thank Marco Moriconi for interesting discussions.
The authors are partially supported by CNPq.

\end{document}